\documentstyle[preprint,aps]{revtex}

\renewcommand{\today}{29 July 1998 \\
Revised: 7 October 1998}

\newcommand{\nc}{\newcommand}
\nc{\be}{\begin{equation}}
\nc{\ee}{\end{equation}}
\nc{\bea}{\begin{eqnarray}}
\nc{\eea}{\end{eqnarray}}
\nc{\beas}{\begin{eqnarray*}}
\nc{\eeas}{\end{eqnarray*}}
\nc{\noi}{\noindent}
\nc{\sD}{\not \! \! D}
\nc{\s}[1]{\not \! #1}
\nc{\non}{\nonumber}
\nc{\bb}{\bibitem}
\nc{\lf}{\left}
\nc{\ri}{\right}
\nc{\mb}[1]{\makebox[#1]{}}
\nc{\pa}{\partial}
\nc{\sA}{\not \! \! A}
\nc{\h}{\frac{1}{2}}
\nc{\ra}{\rightarrow}
\nc{\la}{\leftarrow}
\nc{\ep}{$e^+e^-\ra\pi^+\pi^-\;$}
\nc{\emuon}{$e^+e^-\ra\mu^+\mu^-\;$}
\nc{\epp}{$e^+e^-\ra\pi^+\pi^0\pi^-\;$}
\nc{\elec}{$e^+e^-\ra\gamma^*\ra e^+e^-\;$}
\def\mathunderaccent#1{\let\theaccent#1\mathpalette\putaccentunder}
\def\putaccentunder#1#2{\oalign{$#1#2$\crcr\hidewidth
\vbox to.2ex{\hbox{$#1\theaccent{}$}\vss}\hidewidth}}

\nc{\ti}{\mathunderaccent\widetilde}
\nc{\M}{{\cal M}}
\nc{\rw}{$\rho\!-\!\omega\;$}
\nc{\bold}[1]{\mbox{\boldmath $#1$}}
\nc{\lrpa}{\stackrel{\leftrightarrow}{\pa}}

\begin{document}

\tightenlines
\draft
\preprint{\vbox{\null \hfill  ADP-98-45/T315 \\
                                       \null \hfill LPNHE 98--03 \\
                                       \null \hfill UK/98-04 \\
                                       \null\hfill hep-ph/9807537\\
{\bf Phys.Rev.D59,074020 (1999)}}}
\title{Vector meson dominance and the $\rho$ meson}
\author{M.~Benayoun$^a$, H.B. O'Connell$^b$ and A.G.~Williams$^c$}
\address{
$^a$LPNHE des Universit\'es Paris VI et VII--IN2P3, Paris,
France\\
benayoun@in2p3.fr\\
$^b$Department of Physics and Astronomy, \\ University of Kentucky,
        Lexington, KY 40506-0055, USA\\
        hoc@pa.uky.edu \\
$^c$Department of Physics and Mathematical
Physics \\
and Special Research Centre for the Subatomic Structure of Matter, \\
University of Adelaide 5005,   Australia  \\
  awilliam@physics.adelaide.edu.au}

\date{\today}
\maketitle

\begin{abstract}

We discuss the properties of vector mesons, in particular the $\rho^0$, in
the context of the Hidden Local Symmetry (HLS) model.  This provides a
unified framework to study several aspects of the low energy QCD sector.
Firstly, we show that in the HLS model the physical photon is massless,
without requiring off field diagonalization.  We then demonstrate the
equivalence of HLS and the two existing representations of vector meson
dominance, VMD1 and VMD2, at both tree level and one loop order.  Finally
the S matrix pole position is shown to provide a 
model and process
independent means of specifying the $\rho$ mass and width, in contrast to
the real axis prescription currently used  in the Particle
Data Group tables.

{\it Keywords:
Flavor symmetries,
Chiral symmetries,
Spontaneous symmetry breaking,
Chiral Lagrangians,
Vector-meson dominance,
Analytic properties of S matrix.}

\end{abstract}

\pacs{PACS numbers:
11.30.Hv,
11.30.Rd,
11.30.Qc,
12.39.Fe,
12.40.Vv,
11.55.Bq}

\section{Introduction}

There is no reliable analytic means for calculating low and medium energy
strongly interacting processes with the underlying theory, QCD.  Despite
progress in numerical studies of QCD both via the lattice \cite{lattice}
and QCD--motivated models \cite{DSE}, pre-QCD effective Lagrangians
involving the observed hadronic spectrum continue to play an important role
in studies of this sector.  We shall be concerned in this work with the
interactions of the pseudoscalar and vector mesons as described by the
Hidden Local Symmetry (HLS) model \cite{bando}.

The focus of our paper will be the $\rho$ resonance.  As the lightest and
broadest of the vector octet it plays an important role in phenomenology
and is presently the subject of interest as a possible indicator of chiral
symmetry restoration in heavy ion collisions \cite{BR}.  It also serves as
a guide for physics in other sectors.  As we shall see the interaction of
vector mesons and photons in the HLS construction is closely analogous to
the $SU(2)\otimes U(1)$ symmetry breaking of the electroweak interaction.
The traditional determination of the $\rho$ mass and width has been plagued
by model dependence, which, as we show can be avoided by use of the
S-matrix pole, closely following developments in the study of the $Z^0$.

Our paper is structured as follows:  we begin with a brief outline of the
HLS model and discuss the generation of vector boson masses by the
spontaneous breaking of the global chiral symmetry through the
Higgs--Kibble (HK) mechanism \cite{HK}.  In Sec.~\ref{dress} we consider
the relationship between these HK masses and the {\em physical} vector
masses, using a two-channel propagator matrix for the photon-$\rho$ system.
This allows one to consider the effect of mixing in the dressing of the
bare propagators and when this is properly considered the dressed photon is
seen to be massless as required, without the need for a change of basis.

Sec.~\ref{VvV} is devoted to a comparison of the two commonly used
representations of vector meson dominance (VMD) (referred to hereafter as
VMD1 and VMD2 following the convention of Ref.~\cite{review}), both of
which can be obtained as special cases from the HLS Lagrangian of
Ref.~\cite{bando}.  Note that VMD2 is the most commonly used version and is
often in the literature simply referred to as VMD or ``the vector dominance
model'' \cite{W}.  Using the pion and kaon form-factors we explicitly
demonstrate their equivalence at tree level (which is trivial) and at
one-loop order, where care is required.  This treatment is performed in the
general case $a\neq 2$, where $a$ is the HLS parameter \cite{bando}, for
which we introduce VMD1$_a$ and VMD2$_a$ in an obvious manner.

Finally, as the vector mesons are resonant particles, we study
the effect of the large width on the determination of model independent
$\rho$ parameters. The S-matrix pole position is shown to provide
a truly model-independent and furthermore, process-independent
parametrization
{of the} $\rho$ {meson}.

\section{Hidden Local Symmetry and VMD}
\label{HLS-VMD}

The HLS model
allows us to produce a theory with vector mesons as the gauge
bosons of a hidden local symmetry. These then become massive because
of the spontaneous breaking of a chiral $U(3)_L\otimes U(3)_R$
global symmetry. Let us
consider the chiral Lagrangian \cite{CCWZ},
\be
{\cal L_{\rm chiral}}=\frac{1}{4}{\rm Tr}
[\pa_\mu  F\pa^\mu F^{\dagger}],
\label{ccwz}
\ee
where $F(x)=f_\pi U(x)$ in the usual notation.
This exhibits the chiral $U(3)_L\otimes U(3)_R$ symmetry under
$U\ra g_LUg_R^{\dagger}$.
We can write this in exponential form and expand
\be
F(x)=f_\pi e^{2iP(x)/f_\pi }=f_\pi +2iP(x)-2P^2(x)/f_\pi+\cdots\label{exp}
\ee
therefore, substituting in to Eq.(\ref{ccwz}) we see 
the vacuum corresponds
to $P=0$, $U=1$. That is, $F$ has a non-zero vacuum expectation
value which spontaneously breaks the $U(3)_L\otimes U(3)_R$
symmetry {as this symmetry of the
Lagrangian is not a symmetry of the vacuum}. 
The massless Goldstone bosons contained in $P$,
{ associated with the spontaneous symmetry breaking}
then correspond to the perturbations
about the QCD vacuum and we can think of expansions in this field
given by the hermitian matrix
$P=P^aT^a$ where the Gell-Mann matrices
are normalized such that ${\rm Tr}[T^aT^b]=\delta^{ab}/2$.
So for the pseudoscalars one has
\be
P=\frac{1}{\sqrt{2}}
  \left( \begin{array}{ccc}
            \frac{1}{\sqrt{2}}\pi^0+\frac{1}{\sqrt{6}}\pi_8+
            \frac{1}{\sqrt{3}}\eta_0&\pi^+ &  K^+ \\
            \pi^-  & -\frac{1}{\sqrt{2}}\pi^0+\frac{1}{\sqrt{6}}\pi_8
            +\frac{1}{\sqrt{3}}\eta_0  &  K^0 \\
            K^-             &  \overline{K}^0  &
             -\sqrt{\frac{2}{3}}\pi_8 +\frac{1}{\sqrt{3}}\eta_0 \\
         \end{array} \label{pseudoscalar}
  \right).
\ee

However, in addition to the global chiral symmetry,
$G$, the HLS scheme includes
a local symmetry, $H$, in Eq.(\ref{ccwz}) in the following way
\cite{bando}. 
Let
\be
U(x)\equiv \xi^{\dagger}_L(x)\xi_R(x)\label{f-def}
\ee
where
\bea
\xi_{R,L}(x)&=&\exp[{iS(x)/{f_S}}]\exp[{\pm iP(x)/{f_\pi}}]\label{SP}\\
\xi_{R,L}(x)&\ra&h(x)\xi_{R,L}(x)g^{\dagger}_{R,L}\;.\label{xitransf}
\eea
note that this introduces the scalar field, $S(x)$, analogous to
to the pseudoscalar, $P(x)$ of Eq.~(\ref{pseudoscalar}),
{ though $S(x)$ does not appear in the chiral field $U(x)$
of Eq.~(\ref{f-def}). 
The general forms of the transformations are
given by $g_{L,R}=\exp(i\epsilon_{L,R}^aT^a)$ and
$h(x)=\exp(i\epsilon_{H}^a(x)T^a)$.}

One now seeks to incorporate HLS into the low energy Lagrangian
in a non-trivial way, thereby introducing the lightest vector
meson states \cite{bando,BKYr}.
The procedure is to first re-write ${\cal L_{\rm chiral}}$
explicitly in terms of the $\xi$ components  
\be
{\cal L_{\rm chiral}}=-\frac{f_\pi^2}{4}{\rm Tr}\left[
\pa_\mu\xi_L\xi_L^{\dagger}-\pa_\mu\xi_R\xi_R^{\dagger}\right]^2.
\label{doon}
\ee
The Lagrangian can be gauged for both electromagnetism and the
hidden local symmetry by changing to covariant derivatives
\be
D_\mu\xi_{L,R}=\pa_\mu\xi_{L,R} -igV_\mu\xi_{L,R}+ie\xi_{L,R}A_\mu Q
\ee
where 
$A_\mu$ is the photon field,
$Q=$diag$(2/3,-1/3,-1/3)$ and
$V_\mu=V^a_\mu T^a$ 
where $V^a_\mu$ are the vector meson fields
transforming as 
$V_\mu\longrightarrow h(x)V_\mu h^{\dagger}(x)+{i}h(x)
\pa_\mu h^{\dagger}(x)/{g}$.
Suppressing for brevity the space-time index $\mu$, we can write
the vector meson field matrix $V_\mu$ as
\be
V=\frac{1}{\sqrt{2}}
  \left( \begin{array}{ccc}
   (\rho^0+\omega)/\sqrt{2}  & \rho^+                  &  K^{*+} \\
            \rho^-           &(-\rho^0+\omega)/\sqrt{2}&  K^{*0} \\
            K^{*-}           & \overline{K}^{*0}       &  \phi   \\
         \end{array}\label{vector}
  \right).
\ee
Here we have {used} ideal mixing in defining the 
{bare}
 $\omega$ and $\phi$ mesons.
{Note that the vector meson fields $V^a_\mu=K^*,\rho,\omega,\phi$
are introduced in the role of gauge fields for 
{ $H\equiv$} flavor SU(3).}
However, ${\cal L}_{A}\equiv{\cal L_{\rm chiral}}$ 
is independent of $V$, and 
{in the HLS model}
a second 
{piece of the}
Lagrangian, ${\cal L}_V$, is introduced
\bea\non
{\cal L}_{A}&=&-\frac{f_\pi^2}{4}{\rm Tr}\left[
(D_\mu\xi_L\xi_L^{\dagger}-D_\mu\xi_R\xi_R^{\dagger})\right]^2
\\
{\cal L}_V&=&-\frac{f_\pi^2}{4}{\rm Tr}\left[
(D_\mu\xi_L\xi_L^{\dagger}+D_\mu\xi_R\xi_R^{\dagger})\right]^2.
\label{mass1}
\eea
The full HLS Lagrangian is then
{finally defined}
by  \cite{bando}
\be
{\cal L}_{\rm HLS}={\cal L}_{A}+a{\cal L}_{V} \;, \label{fullHLS}
\ee
{where we see that} 
the HLS parameter $a$
{has now been introduced.}

In the absence of
the gauge fields, $V_\mu$ and $A_\mu$, { we see that Eq.~(\ref{mass1})
reduces to}
\be
{\cal L}_{A}=\frac{1}{2}{\rm Tr}[\pa_\mu P\pa^\mu P],\,\,\,\,\,\,\,\,
{\cal L}_V=\frac{f_\pi^2}{2f_S^2}{\rm Tr}[\pa_\mu S\pa^\mu S].
\label{mass2}
\ee
{ to quadratic order in bosons.} In this case
$P$ and $S$, would both
be Goldstone
bosons, {where $P(x)$ is associated with the spontaneous 
breaking of the usual, global chiral symmetry, $G$,
of ${\cal L}_A$
arising from the vacuum expectation value of the $U(x)$ fields
and analogously for $S(x)$.}
However, gauge invariance allows for their elimination.
It is usual to take a special gauge, the unitary gauge,
for {$H$},
for which the scalar field no longer appears, $S(x)=0$ \cite{BKYr}
(for a discussion of the unitary gauge and spontaneously
broken symmetries, see for example Section 12-5-3 of Ref.~\cite{IZ}).
This is
{phenomenologically}
reasonable as no chiral partner for the pion
{has been observed}.
With this
{choice we have}
\be \xi_L^{\dagger}(x)=\xi_R(x)\equiv\xi(x)=\exp[iP(x)/f_\pi].\ee
By demanding $S(x)=0$
the local symmetry, $H$, is lost, but 
the 
$g$ transformations 
{of the global chiral symmetry group $G$}
will regenerate the scalar field through
\be
\xi(x)\ra\xi'(x)=\xi(x)g^{\dagger}=
\exp[iS'(P(x),g)/f_S]\exp[iP'(x)/f_\pi].
\ee
However the
system {can}
still maintain the 
{unitarity gauge condition $S(x)=0$ and}
global chiral symmetry, $G$, under the combined
transformation \cite{BKYr}
\be
\xi(x)\ra\xi'(x)=h(P(x),g)\xi(x)g^{\dagger},\hspace{1cm}
h(P(x),g)=\exp[-iS'(P(x),g)/f_S]
\ee
where the 
particular choice of
local transformation, $h(P(x),g)$,
``kills" the scalar field created by
the global transformation $g$.
The physical meaning of this is that vector fields acquire
longitudinal components by ``eating" the scalar $S$ field through a
transformation of the form (to lowest order in Goldstone fields)
\be
V_\mu\ra V_\mu-\frac{1}{gf_S}\pa_\mu S'.\label{eat}
\ee

Once the scalar field is 
effectively
removed 
by the unitarity gauge choice
the traditional VMD Lagrangian, that of VMD2,
is obtained from an expansion 
of ${\cal L}_{\rm HLS}$ 
in the pion field, as per Eq.~(\ref{exp}).
The HLS model actually generalizes VMD2, through
the additional parameter, $a$ of Eq.~(\ref{fullHLS}),
so we shall refer to 
the resulting Lagrangian as
VMD2$_a$. 
It has the form
\bea \non
{\cal L}_{{\rm VMD2}a}&=&
-\frac{1}{4}F_{\mu\nu}F^{\mu\nu}-\frac{1}{4}V_{\mu\nu}V^{\mu\nu}
-aef_\pi^2g\left[\rho+\frac{\omega}{3}-
\frac{\sqrt{2}}{3}\phi\right]\cdot A
\\
\non
&+&\frac{2}{3}ae^2f_\pi^2A^2
+\frac{af_\pi^2g^2}{2}(\rho^2+\omega^2+\phi^2)
+af_\pi^2g^2(\rho^+\cdot\rho^-+{K}^{*+}\cdot K^{*-}
+\bar{K}^{*0}\cdot K^{*0})\\ 
\non
&+&\pa\pi^+\cdot\pa\pi^-+\pa K^+\cdot\pa K^-
+\pa K^0\cdot\pa \bar{K}^0+\frac{1}{2}
\left[\pa\pi^0\cdot\pa\pi^0
+\pa\pi^8\cdot\pa\pi^8
+\pa\eta^0\cdot\pa\eta^0\right]
\\ \non
&+&\frac{i}{2}
\left[a g\rho+e(2-a)A\right]\cdot
\left[\pa\pi^+\pi^--\pa\pi^-\pi^+\right]\\
\non
&+&\!\!\frac{i}{4}\left[ag(\rho+\omega-\sqrt{2}\phi )
+2e(2-a)A\right]\cdot(\pa K^+K^--\pa K^-K^+)\\ \non
&+&\!\!\frac{iag}{4}\left[\rho-\omega+\sqrt{2}\phi \right]\cdot
   \left[\pa\overline{K}^0 K^0-\pa{K}^0\overline{K}^0\right]\\ \non
&+&\!\!\frac{iag}{2\sqrt{2}}\rho^+\cdot\left[\pa K^0K^--\pa K^-K^0
+\sqrt{2}(\pa\pi^-\pi^0-\pa\pi^0\pi^-)\right]\\ \non
&+&\!\!\frac{iag}{2\sqrt{2}}\rho^-
\cdot\left[\pa K^+\bar{K}^0-\pa \bar{K}^0K^+
+\sqrt{2}(\pa\pi^0\pi^+-\pa\pi^+\pi^0)
\right]\\ \non
&+&\!\!\frac{iag}{4}K^{*0}\cdot
\left[\pa\pi^0\bar{K}^0-\pa\bar{K}^0\pi^0
+\sqrt{2}(\pa K^-\pi^+-\pa\pi^+ K^-)
+\sqrt{3}(\pa\bar{K}^0\pi^8-\pa\pi^8\bar{K}^0)\right]
\\ \non
&+&\!\!\frac{iag}{4}\bar{K}^{*0}\cdot\left[\pa{K}^0\pi^0-\pa\pi^0{K}^0
+\sqrt{2}(\pa\pi^- K^+-\pa K^+\pi^-)
+\sqrt{3}(\pa\pi^8K^0-\pa K^0\pi^8)\right]
\\ \non
&+&\!\!\frac{iag}{4}{K}^{*-}\cdot\left[\pa\pi^0{K}^+-\pa{K}^+\pi^0
+\sqrt{2}(\pa\pi^+ K^0-\pa K^0\pi^+)
+\sqrt{3}(\pa\pi^8K^+-\pa K^+\pi^8)
\right]
\\
&+&\!\!\frac{iag}{4}K^{*+}\cdot\left[\pa K^-\pi^0-\pa \pi^0 K^-
+\sqrt{2}(\pa\bar{K}^0\pi^--\pa\pi^-\bar{K}^0)
+\sqrt{3}(\pa K^-\pi^8-\pa\pi^8K^-)\right] \non \;,\\
&&\label{lag1}
\eea
where for brevity used the notation
$\rho\equiv\rho^0$. 
$F_{\mu\nu}$ and $V_{\mu\nu}$ are the standard Abelian and
non-Abelian field strength tensors for the photon and vector meson
octet respectively.
The photon and vector mesons therefore acquire Lagrangian masses
through
the Higgs-Kibble (HK) mechanism \cite{HK}
with the vector mesons also obtaining longitudinal components through
Eq.~(\ref{eat}). The HK mass
generated for the vector mesons is given by
\be
m_{HK}^2\equiv M^2=ag^2f_\pi^2.\label{HKmass}
\ee
The choice $a=2$ \cite{bando} reduces VMD2$_a$ (Eq.~(\ref{lag1})) to the
traditional VMD2, eliminating the photon-pion contact term and reproducing the
KSRF relation \cite{KSRF} through Eq.~(\ref{HKmass}).

For simplicity we shall not consider SU(3) breaking due to strangeness in
the vector meson sector \cite{break,BO}, which results in changes in the
vector meson coupling constants and HK masses for the $K^*$ and $\phi$.
One should also note that the VMD2$_a$ Lagrangian does not generate
couplings of the vectors to the {\em singlet} component of the pseudoscalar
mesons.  To be perfectly clear, VMD2$_a$ is nothing but the standard HLS
Lagrangian used to deduce the HLS pion form factor in Ref.~\cite{BEMOSW}.
It is a non--trivial extension of {the traditional vector meson dominance
model} VMD2, since $a \ne 2$ generates not only a violation of the KSRF
relation $(m^2_\rho=2g^2f^2_\pi$), but also introduces a direct coupling of
the photon to the pseudoscalars.

\section{Dressed vector propagators and the photon pole}
\label{dress}
The physical photon is massless, but 
the VMD2$_a$ Lagrangian in
Eq.~(\ref{lag1}) possesses
a photon mass term. To find the physical results one obtains
from VMD2$_a$,
we need to dress the vector propagators
by means of a matrix equation which accounts for the possible mixings
between vector particles \cite{OPTW}.
The matrix Dyson-Schwinger equation is given by
\be
iD_{\mu\nu}=iD^0_{\mu\nu}+iD^0_{\mu\alpha}iG^{\alpha\beta}iD_{\beta\nu}
\label{DSE}
\ee
where $D_{\mu\nu}$ is the dressed propagator matrix and $D^0_{\mu\nu}$
is the bare propagator obtained from the Lagrangian,
\be
D^0_{\mu\nu}=\left(-g_{\mu\nu}+\frac{q_\mu q_\nu}{M^2}\right)
\frac{1}{q^2-M^2}.
\ee
Inverting Eq.~(\ref{DSE}) we obtain
\be
D^{(-1)}_{\mu\nu}=
{D^0}^{(-1)}_{\mu\nu}+G_{\mu\nu}.\label{invert}
\ee

The 
self-energy
matrix $G_{\mu\nu}$ is composed of two entities.
The first is the polarization
function $\Pi_{\mu\nu}$ generated from loop effects.
{As can be shown from Eq.~(\ref{lag1})}, 
the vector mesons couple to
{loops only through}
conserved currents \cite{BO}, and hence the
polarization tensor is transverse
\be
\Pi_{\mu\nu}\equiv(g_{\mu\nu}-{q_\mu q_\nu}/{q^2})\Pi(q^2)
\ee
and from the node theorem \cite{OPTW}
\be \Pi(0)=0.
\ee
The function $\Pi(s)$ has a branch cut
along the real axis beginning at the production threshold (for the
$\rho$ this is $s=4m_\pi^2$) and extending to infinity.
The second dressing term comes from the Lagrangian
mixing terms between the vector mesons and photon. For
simplicity let us consider the case with only the
photon and the  $\rho\equiv\rho^0$ \cite{bando}, we then have
\be
G_{\mu\nu}=\left(g_{\mu\nu}-\frac{q_\mu q_\nu}{q^2}\right)
\left(\begin{array}{cc}
        \Pi_{\gamma\gamma}&\Pi_{\gamma\rho}\\
        \Pi_{\rho\gamma}&\Pi_{\rho\rho}\\
      \end{array}
\right)
+g_{\mu\nu}\frac{eM^2}{g}
\left(\begin{array}{cc}
        0&1\\
        1&0\\
      \end{array}
\right),
\ee
where $M$ is the HK mass of the $\rho$ given in Eq.~(\ref{HKmass}).

As we shall
{here only}
consider amplitudes for such
as \ep where the vectors couple to 
{external}
conserved currents (lepton or pion)
the $q_\mu q_\nu$ pieces of Eq.~(\ref{invert}) can be ignored.
So defining the scalar part of the propagator through 
$D_{\mu\nu}(q)\equiv g_{\mu\nu}D(q^2)$ the 
surviving part of the
dressed propagator is given by
\be
\left(\begin{array}{cc}
        D_{\gamma\gamma}&D_{\gamma\rho}\\
        D_{\rho\gamma}&D_{\rho\rho}\\
      \end{array}
\right)=\left(\begin{array}{cc}
e^2M^2/g^2+\Pi_{\gamma\gamma}-s\,\,\,\,\,\,&eM^2/g+\Pi_{\gamma\rho}\\
        eM^2/g+\Pi_{\rho\gamma}&M^2+\Pi_{\rho\rho}-s\\
      \end{array}
\right)^{-1},
\ee
where $s\equiv q^2$.
A similar formalism has been
discussed by Hung and Sakurai for the $\gamma-Z^0$
system in the electroweak interaction \cite{HS}. The poles, $p_i$,
$i=1,2$, of the
two channel propagator are obtained from
\be
{\rm Det}[D^{-1}(p_i)]=0.\label{det-eq}
\ee
The 
{VMD2$_a$}
Lagrangian
{given in}
Eq.~(\ref{lag1}) is ${\cal O}(\epsilon^2)$,
where $\epsilon\equiv e/g$, so we
work to this order in the calculation of the pole positions in the
complex $s$--plane. Eq.~(\ref{det-eq}) gives
\bea\non
2p&=&M^2(1+\epsilon^2)+\Pi_{\rho\rho}+\Pi_{\gamma\gamma}\\
&&\pm\left\{M^2(1+\epsilon^2)+\Pi_{\rho\rho}-\Pi_{\gamma\gamma}
+\frac{2(\epsilon^2(\Pi_{\gamma\gamma}-\Pi_{\rho\rho})M^2+2\epsilon
M^2\Pi_{\gamma\rho}-\Pi_{\gamma\rho}^2)}{M^2(1+\epsilon^2)
+\Pi_{\rho\rho}-\Pi_{\gamma\gamma}}+{\cal O}(\epsilon^3)\right\}.
\eea
In power counting with respect to $\Pi_{\rho\rho}$,  we see that
$\Pi_{\gamma\rho}$ and $\Pi_{\gamma\gamma}$
are intrinsically of ${\cal O}(\epsilon)$ and ${\cal O}(\epsilon^2)$
respectively, so we can further simplify,
\bea\non
2p&=&M^2(1+\epsilon^2)+\Pi_{\rho\rho}+\Pi_{\gamma\gamma}\\
&&\pm\left\{M^2(1+\epsilon^2)+\Pi_{\rho\rho}-\Pi_{\gamma\gamma}
-\frac{2(\epsilon^2M^2\Pi_{\rho\rho}-2\epsilon
M^2\Pi_{\gamma\rho}+\Pi_{\gamma\rho}^2)}{M^2
+\Pi_{\rho\rho}}+{\cal O}(\epsilon^3)\right\}.
\eea
Thus the poles are, to ${\cal O}(\epsilon^2)$,
\bea
p_\gamma&=&\Pi_{\gamma\gamma}(p_\gamma)+
\frac{\epsilon^2M^2\Pi_{\rho\rho}(p_\gamma)-2\epsilon
M^2\Pi_{\gamma\rho}(p_\gamma)+\Pi_{\gamma\rho}^2(p_\gamma)}{M^2
+\Pi_{\rho\rho}(p_\gamma)}\\
p_\rho&=&M^2(1+\epsilon^2)+\Pi_{\rho\rho}(p_\rho)-
\frac{\epsilon^2M^2\Pi_{\rho\rho}(p_\rho)-2\epsilon
M^2\Pi_{\gamma\rho}(p_\rho)+\Pi_{\gamma\rho}^2(p_\rho)}{M^2
+\Pi_{\rho\rho}(p_\rho)}.
\eea
{Since they couple only through conserved currents,} 
the polarization functions, $\Pi(0)=0 \cite{OPTW}$, and
{we see that}
the photon pole resides at $q^2=0$ as required. 
{The explicit photon mass term in Eq.~(\ref{lag1})
is canceled by the photon self-energy.}
Similarly the $\rho$ pole is shifted
only by corrections of ${\cal O}(\epsilon^2)$ by mixing with the
photon.  
This result is as we might expect, since the vacuum state $F_0=f_\pi$
in invariant under the U(1) EM transformation \cite{witten}
\be
F\ra F+i\epsilon(x)[Q,F],
\ee
and massive vectors correspond to spontaneously broken symmetries.
However, it is slightly more subtle than the usual case where invariance
under a transformation of the form 
$F_0\ra F_0+i\epsilon^a(x)T^aF_0$ requires that
there be no Lagrangian mass term (as $T^aF_0=0$).
The generation of vector meson masses in the
presence of a massless photon is treated in the general case
by Gottlieb \cite{G}.

HLS can therefore be used to demonstrate ``from first principles"
that the VMD2 Lagrangian is perfectly consistent with
physical expectations concerning the
photon and vector meson
masses. However, this is only realized 
after the above cancellation of the photon mass.
For this reason, an alternative representation of
vector meson dominance, VMD1, was introduced. In the next section we 
examine
the relationship between VMD1 and VMD2 at both tree level and 
one-loop
order.

\section{VMD1 versus VMD2}
\label{VvV}
The VMD2$_a$ Lagrangian, Eq.~(\ref{lag1}), contains
a mass term for the photon.
As
the physical photon is massless, one might consider
an alternative version
with no Lagrangian mass term for the photon \cite{KLZ},
which
has been
referred to as VMD1~\cite{review,OPTW2}.
We shall introduce the term VMD1$_a$ for the general version of
VMD1 derived from HLS. As in the previous representation of VMD, VMD1$_a$
reduces to standard VMD1 for 
the special case
$a=2$.

In VMD1$_a$ the photon can couple directly to the pseudoscalars and the
photon--vector meson mixing term is linear in $q^2$ and hence vanishes at
$q^2=0$.  As will be shown the two representations of VMD, VMD1$_a$ and
VMD2$_a$, are related by a field transformation, and as such should be
physically equivalent.  At tree level, this equivalence is complete and
easy to prove, however, at one loop level care needs to be taken to ensure
all terms are included \cite{KKW}.  We begin with a discussion of one-loop
effects in VMD, starting from the HLS Lagrangian.

\subsection{VMD2$_a$ form-factors}

So far we have discussed the generation of masses 
and longitudinal components
for the vector mesons through the HK mechanism.
These masses are necessarily real-valued, but
are not exactly what is seen in
experiment. The vector meson propagator
is modified by the vacuum polarization
function $\Pi(q^2)$
{away from $q^2=0$, i.e.},
\be
D_{V}(s)=\frac{1}{s-m_{HK}^2}\ra\frac{1}{s-m_{HK}^2-\Pi_{VV}(s)},
\hspace{1cm}s\equiv q^2.
\ee

As the physical pseudoscalar
fields appear as vacuum fluctuations we may assume a weak field expansion
and work to first order in pseudoscalar loops \cite{BKYr}.
The polarization functions, $\Pi_{VV}(s)$, are composed
of loops, $\ell(PP')$, from the 
$VPP^\prime$ 
couplings
\bea
\Pi_{\rho\rho}&=&{g^2a^2}\ell(\pi^+\pi^-)/4+
{g^2a^2}\ell(K^+K^-)/16+{g^2a^2}\ell(K^0\bar{K}^0)/16 \\
\Pi_{\omega\omega}&=&g^2a^2\ell(K^+K^-)/16+g^2a^2\ell(K^0\bar{K}^0)/16\\
\Pi_{{K^*}^0{K^*}^0}&=&{g^2a^2}\ell(K^-\pi^+)/2+{g^2a^2}
\ell({K}^0\pi^0)/4\\
\Pi_{{K^*}^+{K^*}^+}&=&{g^2a^2}\ell(\bar{K}^0\pi^+)/2+{g^2a^2}
\ell({K}^+\pi^0)/4\\
\Pi_{\phi\phi}&=&g^2a^2\ell(K^+K^-)/8+g^2a^2\ell(\bar{K}^0K^0)/8
\eea
where we have factored out the couplings leaving only the generic
loop integrals involving the appropriate pseudoscalars.
Note that we have not included the 
anomalous 
$VPPP$ and $VVP$ couplings \cite{anom}
in the polarization functions,
since these effects are expected to be rather small.
Numerical studies show that the
$\rho\ra\pi\pi\ra\rho$ contribution to the real part of $\Pi_{\rho\rho}$
is a few percent of 
$m_{HK}\equiv M$, but the contribution to
the $\omega$ physical mass from $\omega\ra3\pi\ra\omega$ is negligible
\cite{numer}. Similar behavior is observed in the imaginary part,
as the $\rho$ width (generated by Im$\Pi_{\rho\rho}$) is
{\em much} larger
than that of the $\omega$ or $\phi$.
We may conclude that the two pion loop is
the dominant influence\footnote{See, however, Ref.~\cite{KKW}
for a discussion of the $\phi$ meson.}
on the mass shift and that only the $\rho$
physical mass is significantly different from its HK value. 
This is supported
by the observation that
the Bando-Kugo-Yamawaki relation between the
HK masses of the $\omega$, $K^*$ and $\phi$
in models of the SU(3) breaking \cite{break},
\be
m_\omega m_\phi=m_{K^*}^2
\ee
holds to $0.1-0.4\%$ (for charged or neutral $K^*$), whereas the observed
$\rho$ and $\omega$ masses differ by a few percent despite 
having identical
HK masses.

In a similar manner, we
can obtain the hadronic contributions loop corrections to vector meson
mixing, and can derive relations between them and the polarization 
functions. The pure
pseudoscalar loop mixings are
given by
\bea\label{rwmix}
\Pi_{\rho\omega}&=&\frac{g^2a^2}{16}[\ell(K^+K^-)-\ell(K^0\bar{K}^0)] \\
\Pi_{\rho\phi}&=&-\frac{\sqrt{2}g^2a^2}{16}
[\ell(K^+K^-)-\ell(K^0\bar{K}^0)]\label{rpmix} \\
\Pi_{\phi\omega}&=&-\frac{\sqrt{2}g^2a^2}{16}[\ell(K^+K^-)+
\ell(K^0\bar{K}^0)].
\label{pwmix}
\eea
Note that if
isospin invariance
is assumed, as it has been so far, that
only $\omega\!-\!\phi$ mixing survives and we notice that
\be
\Pi_{\phi\phi}=2\Pi_{\omega\omega}=-2\sqrt{2}\Pi_{\phi\omega}.
\label{pols}
\ee
Hence $\rho\!-\!\omega$ mixing is only allowed in the present
analysis if isospin violation is admitted through allowing
$\ell(K^+K^-)\neq\ell(K^0\bar{K}^0)$.  However, once isospin violation
is allowed, we should in principle 
consider the possibility of
additional isospin violating 
effects arising from
$u-d$ splitting in our HLS Lagrangian.
For example, to first order in isospin violation the possibility
of an $\omega\pi\pi$ coupling must be considered in
$\rho\!-\!\omega$ mixing \cite{mix_int,GO}. Such a term
arises naturally via loop effects when higher order pseudoscalar terms,
for example $VPPPP$, are considered \cite{chiral}.
Alternatively it could be generated through SU(2) 
breaking in ${\cal L}_V$ 
itself, analogous to the existing studies of SU(3) breaking 
\cite{break,BO}. So far, in studies of the
HLS model such explicit isospin breaking effects 
have been considered only for the anomalous sector 
\cite{iso-anom,iso-anom2}. For an investigation of isospin violation
in a chiral meson theory see Ref.~\cite{iso-na}.

The possibility of a direct coupling of the photon to the pseudoscalar
field allows for a loop-induced photon--vector meson mixing, which in the
isospin limit (where $m_{K^0}=m_{K^+}$) gives
\be
\Pi_{\rho\gamma}=\frac{e(2-a)}{ag}\Pi_{\rho\rho},\,\,\,\,\,\,\,\,
\Pi_{\omega\gamma}=\frac{e(2-a)}{ag}\Pi_{\omega\omega},\,\,\,\,\,\,\,\,
\Pi_{\phi\gamma}=-\frac{e(2-a)}{\sqrt{2}ag}\Pi_{\phi\phi}.\label{mix2}
\ee
 Note that these mixings vanish if $a=2$ or $q^2=0$.  Since
these vanish at $q^2=0$, we see that the previous 
proof of the masslessness
of the photon remains unaffected.

{From} the Lagrangian we can now write expressions for the pion and
kaon form-factors, defined from the Feynman amplitude
via ${\cal M}_{\gamma PP}=-eF_{P}$.
In the following we shall write the HK mass simply as $M$
(see Eq.~(\ref{HKmass})).
We can go from the tree level result to the one-loop result by replacing
the ordinary Lagrangian interaction pieces, by an effective Lagrangian
for the photon pseudoscalar couplings.
At tree-level, where the propagators are simply $D_V^0=1/(s-M^2)$,
the form-factors are given by
\bea
F_\pi^{\rm tree}(s)&=&
1-a/2-aM^2D_\rho^0/2=1-asD_\rho^0/2\label{p+}
\\ \label{K+}
F_{K^+}^{\rm tree}
(s)&=&1-a/2-aM^2[D_\rho^0+D_\omega^0/3+2D_\phi^0/3]/4
=1-(as/4)[D_\rho^0+D_\omega^0/3+2D_\phi^0/3]
\\
F_{K^0}^{\rm tree}
(s)&=&-(aM^2/4)[D_\rho^0-D_\omega^0/3-2D_\phi^0/3].\label{K0}
\eea
We note here that in this SU(3) flavor symmetric tree result
$F_{K^0}^{\rm tree}(s)=0$.

We now consider the effects of loops. As the resonant 
structure of the vector
mesons, generated by loops, is an important part of the 
phenomenology
we might expect loops to play an important role.
The contact term between the photon and the pseudoscalars 
 when $a\neq 2$
induces
a one loop contribution to the photon--vector meson vertex, which we
can introduce through the effective interaction Lagrangian
\bea\non
{\cal L}^{\rm eff}_{\gamma\rho}&=&-\frac{eM^2}{g}+\Pi_{\gamma\rho}
\,\,\,\,\,\,\,=-\frac{eM^2}{g}
+\frac{e(2-a)}{ag}\Pi_{\rho\rho}\\
{\cal L}^{\rm eff}_{\gamma\omega}&=&-\frac{eM^2}{3g}+\Pi_{\gamma\omega}
\,\,\,\,\,\,=-\frac{eM^2}{3g}+
\frac{e(2-a)}{ag}\Pi_{\omega\omega}\label{efflag}\\
{\cal L}^{\rm eff}_{\gamma\phi}&=&\frac{\sqrt{2}eM^2}{3g}
+\Pi_{\gamma\phi}
\,\,\,=\frac{\sqrt{2}eM^2}{3g}
-\frac{e(2-a)}{\sqrt{2}ag}\Pi_{\phi\phi}.\non
\eea
 In a similar way the tree-level propagators are replaced by
their one-loop forms
to give
$D^1_V=1/(s-M_V^2-\Pi_{VV}(s))$. Similarly
the vector mesons can now mix through pseudoscalar loops.
For example,
the effect of \rw mixing,
which is important for the pion form factor,
is easily
incorporated by replacing the $\rho$ propagators by
\be
D^1_\rho\ra D^1_\rho[1+(f_{\omega\gamma}/f_{\rho\gamma})
A_{\omega\pi\pi}D^1_\omega].\label{rhoprop}
\ee
As it is not realistic, in the sense of data fitting \cite{mix_int}, to 
draw the distinction between 
isospin violation occurring through \rw mixing and intrinsic isospin
violation in the $\omega\pi\pi$ vertex (which could 
could either be present in the original Lagrangian or
be generated by
loop effects \cite{chiral}), 
Eq.~(\ref{rhoprop}) uses
$A_{\omega\pi\pi}$, the ``effective mixing function"
that absorbs both effects to couple the $\omega$ to the
2 pion final state, through the $\rho$\cite{mix_int,GO},
\be
A_{\omega\pi\pi}=-3500\pm300\;{\rm MeV}^2.
\ee
Using Eqs. (\ref{pols}) and (\ref{efflag})
{the leading resonant terms in the one-loop expressions for the 
VMD2$_a$ model form-factors
are given by}
\bea
F_\pi^{\rm 1\;loop}(s)&=&[(1-a/2)s-M^2]D^1_\rho
[1+(f_{\omega\gamma}/f_{\rho\gamma}){A_{\omega\pi\pi}}D^1_\omega]
\label{2pi}
\\
F_{K^+}^{\rm 1\;loop}(s)&=&(1/2)[(1-a/2)s-M^2]
[D^1_\rho+(1/3)D^1_\omega+(2/3)D^1_\phi
-(2\sqrt{2}/3)D^1_\omega \Pi_{\omega\phi}D^1_\phi]\label{2kp}\\
F_{K^0}^{\rm 1\;loop}(s)&=&(1/2)[(1-a/2)s-M^2]
[D^1_\rho-(1/3)D^1_\omega-(2/3)D^1_\phi
+(2\sqrt{2}/3)D^1_\omega \Pi_{\omega\phi}D^1_\phi].\label{2k0}
\eea
These expressions are one--loop in the sense that each of the
components in the tree-level amplitudes have been replaced by its
one--loop generalization.
Without
the vector meson mixing
loops, these expressions agree identically with the tree level
results for $a=2$ (as the photon decouples from the
pseudoscalars and pseudoscalar
loops cannot be generated). 
We see is that the
tree level results are protected from major corrections at 
the one-loop level
for $a\simeq 2$, since vector meson mixing effects are small
\cite{HY}.

\subsection{VMD1$_a$ Form Factors}
Although the photon mass term appearing in Eq.~(\ref{lag1}) does not
result in a massive photon, it is tempting to choose a 
field redefinition 
in which it does not appear \cite{bando}. The transformation most
resembling that used to eliminate the photon mass term in the 
electro-weak
theory actually removes the pointlike coupling
of the photon to the $\rho$ and thus
obscures VMD \cite{bando,s1}. We shall consider an alternative 
transformation
\cite{KLZ} that generates ``current mixing" \cite{CS} between 
the photon and 
vector mesons,
\be
A\ra %A'=
A,\,\,\,\rho\ra%\rho'=
\rho+\epsilon A
,\,\,\,\omega\ra%\omega'=
\omega+(\epsilon/3) A
,\,\,\,\phi\ra%\phi'=
\phi-(\epsilon\sqrt{2}/3) A.\label{transf}
\ee
to derive VMD1$_a$ from VMD2$_a$ (Eq.~(\ref{lag1})).
Making the substitutions of Eq.~(\ref{transf}) in Eq.~(\ref{lag1})
we find 
the resulting Lagrangian for VMD1$_a$ is given by
\bea \non
{\cal L}_{{\rm VMD1}a}&=&
-\frac{1}{4}F_{\mu\nu}F^{\mu\nu}-\frac{1}{4}V_{\mu\nu}V^{\mu\nu}
-\epsilon s\left[\rho+\frac{\omega}{3}-\frac{\sqrt{2}}{3}\phi\right]
\cdot A
\\
\non
&+&\frac{1}{2}af_\pi^2g^2(\rho^2+\omega^2+\phi^2)
+af_\pi^2g^2(\rho^+\cdot\rho^-+{K}^{*+}\cdot K^{*-}+
\bar{K}^{*0}\cdot K^{*0})\\ 
\non
&+&\pa\pi^+\cdot\pa\pi^-+\pa K^+\cdot\pa K^-
+\pa K^0\cdot\pa \bar{K}^0+\frac{1}{2}
\left[\pa\pi^0\cdot\pa\pi^0
+\pa\pi^8\cdot\pa\pi^8
+\pa\eta^0\cdot\pa\eta^0\right]
\\ \non
&+&\frac{i}{2}
[a g\rho+2eA]\cdot\left[\pa\pi^+\pi^--\pa\pi^-\pi^+\right]\\
\non
&+&\!\!\frac{i}{4}\left[ag(\rho+\omega-\sqrt{2}\phi )
+4eA\right]\cdot(\pa K^+K^--\pa K^-K^+)\\ \non
&+&\!\!\frac{iag}{4}\left[\rho-\omega+\sqrt{2}\phi \right]\cdot
   \left[\pa\overline{K}^0 K^0-\pa{K}^0\overline{K}^0\right]\\ \non
&+&\!\!\frac{iag}{2\sqrt{2}}\rho^+\cdot\left[\pa K^0K^--\pa K^-K^0
+\sqrt{2}(\pa\pi^-\pi^0-\pa\pi^0\pi^-)\right]\\ \non
&+&\!\!\frac{iag}{2\sqrt{2}}\rho^-\cdot
\left[\pa K^+\bar{K}^0-\pa \bar{K}^0K^+
+\sqrt{2}(\pa\pi^0\pi^+-\pa\pi^+\pi^0)
\right]\\ \non
&+&\!\!\frac{iag}{4}K^{*0}\cdot\left[\pa\pi^0\bar{K}^0-\pa\bar{K}^0\pi^0
+\sqrt{2}(\pa K^-\pi^+-\pa\pi^+ K^-)
+\sqrt{3}(\pa\bar{K}^0\pi^8-\pa\pi^8\bar{K}^0)\right]
\\ \non
&+&\!\!\frac{iag}{4}\bar{K}^{*0}\cdot\left[\pa{K}^0\pi^0-\pa\pi^0{K}^0
+\sqrt{2}(\pa\pi^- K^+-\pa K^+\pi^-)
+\sqrt{3}(\pa\pi^8K^0-\pa K^0\pi^8)\right]
\\ \non
&+&\!\!\frac{iag}{4}{K}^{*-}\cdot\left[\pa\pi^0{K}^+-\pa{K}^+\pi^0
+\sqrt{2}(\pa\pi^+ K^0-\pa K^0\pi^+)
+\sqrt{3}(\pa\pi^8K^+-\pa K^+\pi^8)
\right]
\\
&+&\!\!\frac{iag}{4}K^{*+}\cdot\left[\pa K^-\pi^0-\pa \pi^0 K^-
+\sqrt{2}(\pa\bar{K}^0\pi^--\pa\pi^-\bar{K}^0)
+\sqrt{3}(\pa K^-\pi^8-\pa\pi^8K^-)\right] \non \;,\\
\label{VMD1a}
\eea

This removes the photon mass term as well as the constant $V.A$ terms
from Eq.~(\ref{lag1}) 
and introduces a new coupling of the photon to the pseudoscalar fields
\be
{\cal L}^{\rm VMD1}_{\gamma P}
=ieA^\mu[(\pa_\mu\pi^+\pi^--\pa_\mu\pi^-\pi^+)+
(\pa_\mu K^+K^--\pa_\mu K^-K^+)].\label{VPP}
\ee
The field redefinition in
Eq.~(\ref{transf}) 
also introduces the 
$q^2=0$ vanishing
mixing of the photon and vector mesons through
the vector meson kinetic terms (this is discussed in detail in
Ref.~\cite{review}). Together with Eq.~(\ref{VPP})
we then have the effective interaction Lagrangian
\be
{\cal L}_{\gamma V}^{{\rm VMD1}_a}%&=&
=\frac{e}{g}\left[\left(-{s}+\frac{2}{a}\Pi_{\rho\rho}\right)\rho_\mu
+\left(-\frac{s}{3}+\frac{2}{a}\Pi_{\omega\omega}\right)\omega_\mu
+\left(\frac{\sqrt{2}s}{3}-\frac{\sqrt{2}}{a}\Pi_{\phi\phi}\right)
\phi_\mu
\right]A^{\mu}
\label{vmd1eff}
\ee

We are now in a position to write the VMD1$_a$ form-factors, as derived
from the HLS model.
The tree level result is
\bea
F_\pi^{\rm tree}(s)&=&1-asD^0_\rho
/2\label{1pi}\\
F_{K^+}^{\rm tree}(s)&=&1-(as/4)
[D^0_\rho+D^0_\omega/3+2D^0_\phi/3]\label{1kp}
\\
F_{K^0}^{\rm tree}(s)&=&-(as/4)
[D^0_\rho-D^0_\omega/3-2D^0_\phi/3]\label{1k0}
\eea
We see that Eqs.~(\ref{p+}$-$\ref{K0}) and Eqs.~(\ref{1pi}$-$\ref{1k0})
are identical (recalling Eqs.~(\ref{2k0}) and similarly (\ref{1k0}) are 
zero) and that the two representations of VMD are thus equivalent.
While the 
leading resonant terms for the
one-loop result are given by
\bea
F_\pi^{\rm 1\;loop}(s)&=&[(1-a/2)s-M^2]D^1_\rho
[1+(f_{\omega\gamma}/f_{\rho\gamma}){A_{\omega\pi\pi}}D^1_\omega]
\\
F_{K^+}^{\rm 1\;loop}(s)&=&
(1/2)[(1-a/2)s-M^2][D^1_\rho+(1/3)D^1_\omega
+(2/3)D^1_\phi-(2\sqrt{2}/3)D^1_\omega\Pi_{\omega\phi}D^1_\rho]\\
F_{K^0}^{\rm 1\;loop}(s)&=&(1/2)[(1-a/2)s-M^2][D^1_\rho-(1/3)D^1_\omega
-(2/3)D^1_\phi+(2\sqrt{2}/3)D^1_\omega\Pi_{\omega\phi}D^1_\rho],
\eea
which are identical to Eqs.(\ref{2pi}$-$\ref{2k0})

It is worthwhile now to briefly
explain how the one loop results have been
obtained. The
simplest case is the pion form-factor of VMD1$_a$. {From}
the effective Lagrangian one has (ignoring, for simplicity,
the isospin violating piece)
\be
F_\pi^{\rm 1\;loop}(s)=
1-(as/2-\Pi_{\rho\rho}(s))D_\rho^1.\label{chic}
\ee
The procedure we have followed
is to eliminate the $\Pi_{\rho\rho}(s)$ term appearing
in the numerator of Eq.~(\ref{chic}). Doing this also cancels the leading
1 associated with the pointlike coupling of the photon to the
pion current,
\be
F_\pi^{\rm 1\;loop}(s)=[(s-M^2-\Pi(s))-(as/2-\Pi(s))]D_\rho^1=
[s(1-a/2)-M^2]D_\rho^1.
\ee
The other expressions are obtained similarly. Recalling
$\Pi(0)=0$ \cite{OPTW} all form-factor normalization conditions are 
clearly maintained ($A_{\omega\pi\pi}$ in our model is generated only by 
loops).
We can now clearly see the agreement
between the one loop results for VMD1$_a$ and
VMD2$_a$. This highlights the importance
including loop effects in VMD studies.

So we see the equivalence of VMD1$_a$ and VMD2$_a$, which we might suspect
on principle as the physics must be independent of the choice of fields.
However, this is only true if one works consistently to one loop order in
the form-factors, including loop induced photon-vector meson mixings.  In
addition to this, care must be taken to distinguish between the HK and the
physical mass of the $\rho$.  Such effects were not included in
Ref.~\cite{BEMOSW} leading to somewhat different results for data fits
using VMD1$_a$ and VMD2$_a$ (note that in this paper the ``HLS" fits refer
to what we call VMD2$_a$ here).  For the kaon form-factors it would be
interesting to more systematically include and study the effects of SU(3)
symmetry breaking on this analysis \cite{break}.  One should note, however,
that including SU(3) symmetry breaking effects will not spoil the effect
of the transformation, given in Eq.~(\ref{transf}), on the VMD Lagrangian.

It should also be noted that the initial investigation into generalizations
of VMD1 and VMD2 performed in Ref.~\cite{BEMOSW} can be interpreted as
evidence for preferring VMD1$_a$/VMD2$_a$ to VMD1/VMD2.  Indeed, in order
to get an acceptable fit to the form factor data, it was necessary to
introduce an additional term $(1+\epsilon)=1.167\pm0.008$ affecting the
resonance contribution in the standard VMD1 form factor and interpret this
as an {\it ad hoc} universality violation\footnote{Here $\epsilon$ is
merely a fitting parameter, not $e/g$ as defined earlier.}.  We might, in
light of the above discussion make a connection between $(1+\epsilon)$ and
the HLS $a/2$.  Similarly, the ``HLS" fit of Ref.~\cite{BEMOSW}, i.e., a
VMD2$_a$ type fit finds significant excursion from the traditional $a=2$,
namely, $a/2=1.182\pm 0.008$.  One should then remark that the results of
Ref.~\cite{BEMOSW} give additional evidence for the underlying equivalence
of VMD1$_a$ and VMD2$_a$.  It would be very interesting to refit the data
with the full 1-loop expressions for the form factors given here.

\section{Masses, widths and S matrix poles}

Because
$m_{\rm HK}$ is above threshold,
the vector meson pole, $p_V$, ``slips down" onto the second
Riemann sheet to a location on the complex plane satisfying
\be
p_V-m_{HK}^2-\Pi_{VV}(p_V)=0.\label{poledef}
\ee
In some sense the discussion of masses and widths for unstable particles
is  an artifact of convention, because only the pole position,
$p_V$, defined
in Eq.~(\ref{poledef}) is 
process--independent and hence
physically meaningful \cite{LM}.
Model dependence is, of course, unavoidable when
comparing 
various
resonance models and their different associated vector meson masses
and widths.

However,
model-independent definitions of the
masses and widths 
can be conveniently defined from the pole
position through the identification
\be
p_V\equiv \overline{m}^2-i\overline{m}\overline{\Gamma}\label{pole_def}
\ee
(or through the slightly old-fashioned form
$\sqrt{p_V}\equiv \widetilde{m}-i\widetilde{\Gamma}/2$).
For narrow resonances such as
the $\omega$ and $\phi$, the pole is directly accessed by the choice of
a momentum independent Breit-Wigner
form in the fit to data
and the masses and widths generally quoted correspond 
very closely to those in Eq.~(\ref{pole_def}). However, due
to the large width of the $\rho$ one generally sees an attempt to
model the possible $s$ dependence of $\Pi_{VV}$ modifying the
naive Breit Wigner
(see, e.g., Refs.~\cite{ffs,HL}). For the pion form-factor,
where $\rho$ parameters are extracted,
this amounts to the isospin pure contribution (that not including the
$\omega$ \cite{GO}) being  given by a function
\be
F_\pi(s)=\frac{f(0)}{f(s)}h(s),\hspace{1cm}h(0)=1\label{general}
\ee
where $f(s)$ has the appropriate cut beginning at $s=4m_\pi^2$ and
the function $h(s)$ models any small deviations from elasticity.
Though this in itself is perfectly
acceptable, the quoted values of mass and width are defined through
their values on the real axis
\cite{HL,barkov}
\be
{\rm Re}\:f(m_{\rm Re}^2)=0,\hspace{0.5cm}{\rm Im}
f(m_{\rm Re}^2)=-im_{\rm Re}\Gamma_{\rm Re},\,\,\,\,\,\,\,
m_{\rm Re},\Gamma_{\rm Re}\in\Re
\label{usual}
\ee
to emulate the Breit-Wigner form-factor
along the real axis \cite{PR},
\be
\lim_{s\ra m^2_{\rm Re}}f(s)=s-m^2_{\rm Re}+im_{\rm Re}
\Gamma_{\rm Re}.
\ee
These are not tied to the pole position and are 
hence very model dependent.
Indeed, phenomenology 
suggests
specific forms for the function
$f(s)$, which in turn influence
the parameter values for the function $f(s)$.
This has led to considerable difficulties in
comparing $\rho$ parameters obtained by various authors
as has been noted by the PDG as early as 1971 
(see pg.~S62 of Ref.~\cite{PDG71}) and is also discussed in the most
recent Particle Data listing
(see Eidelman's review on pg.~364 of Ref.~\cite{PDG}). 
Indeed the differences between many quoted values for
the $\rho$ mass and width arise merely from model-dependence,
rather than discrepancies in physical data,
as discussed by Gardner and O'Connell \cite{GO}
for the \ep data of Barkov {\it et al.}~\cite{barkov}.

This is a well known problem for the parametrization of broad resonances.
The model dependence of the determination of the mass and width of the
$\Delta(1236)$ was discussed in the 1971 Particle Data listing
\cite{PDG71}.  The solution, through the model independent pole
prescription, was then provided in the 1972 listing \cite{PDG72}.
H\"{o}hler's recent review of this (see pg.~624 of Ref.~\cite{PDG})
concludes that in contrast to the conventional real axis parameters, the
pole positions have a well-defined relation to $S$ matrix theory and are
generally more useful.  Our aim here is to present the pole results for
various recent fits to $\rho$ data in one place to fully highlight both the
model independence and process independence.

In Ref.~\cite{GO} generalized form-factors were used to obtain
very good fits to the pion form-factor data \cite{barkov}
for four choices of model (A, B, C and D). 
In this manner  the function $f(s)$ is fixed, 
in terms of fitting parameters, along the real $s$ axis.
The resulting real axis determinations of the mass and width
showed a significant model dependence. However, in Ref.~\cite{GO2}
the model independence of the pole prescription 
can clearly be seen in the quoted values of 
$\overline{m}_\rho$ and
$\overline{\Gamma}_\rho$ as defined in Eq.~(\ref{pole_def}) 
by solving
$f(p_V)=0$ for each of the four models.
To give an example this procedure, model A of Ref.~\cite{GO} uses 
$f(s)=as^2 +bs +c+f_1(s)$, where $f_1(s)$ is
a specified {\em complex} function. 
The parameters $a,b,c$, are fixed by the fit to data,
allowing one to find the zero of $f(s)$ 
through a step iteration on the complex plane. 
{The} correlations
associated with the parameters $a$, $b$ and $c$ 
{allow one to} determine the statistical errors associated 
with $\overline{m}_\rho$ and
$\overline{\Gamma}_\rho$ (for a discussion
of error analysis see, for example, Ref.~\cite{stats}).
The results are given in
Table \ref{one} along with the masses and widths obtained from
the traditional real-axis prescription (which show a large
model dependence).

We have also repeated this procedure for other determinations of the $\rho$
mass and width, although without knowing correlations of the fitting
parameters we have simply solved $f(p_V)=0$ using Mathematica \cite{math}
and do not quote statistical errors for $\overline{m}_\rho$ and
$\overline{\Gamma}_\rho$.  This is done for the fits of Ref.~\cite{BEMOSW}
as well as the original fit of Barkov {\it et al.}  \cite{barkov} and the
recent fit for the {\em charged} $\rho$ seen in hadronic $\tau$ decay
\cite{aleph}.  The latter results should allow one to look for any
statistically significant isospin violation in the pole position.  As
$m_{\rho(770)^0}-m_{\rho(770)^\pm}$ is quoted by the PDG \cite{PDG}, a more
detailed study would be of considerable interest for a determination of
this mass splitting based on model-independent quantities.

Not only are the results of the pole determination virtually
model-independent, they agree completely with previous pole determinations
from both \ep \cite{BCP} {\em and} $\pi-\pi$ \cite{scatt1,scatt2}
scattering data (further demonstrating process independence).  We see in
Table~\ref{one} that the central values for the $\rho$ mass and width,
defined by the standard real-axis prescription, cover ranges of $763-780$
and $141-157$ respectively, while those from the pole prescription, lie in
the ranges $756-759$ and $140-145$ respectively (all figures quoted in
MeV).  This illustrates the model independence of the $\rho$ pole location.
Indeed, if one takes into account the errors on the pole parameters, when
they are available, the difference between the various determinations of
the pole position is never greater than $2 \sigma$, i.e., not
statistically significant.

The power of this method is useful not only to finding a model
independent means of parametrizing a broad resonance such
as the $\rho$ mesons, but also in the
Standard Model for the $Z^0$  where the data has
very high precision and $\Pi_{VV}(s)$ can actually be theoretically
calculated \cite{Zpole,sirlin}
(see also Caso's $Z^0$ review in Ref.~\cite{PDG}). 
The use of S-matrix poles is also potentially
important for the Higgs boson \cite{KS} and CP violation in the kaon
system \cite{kaon}.

\section{Conclusion}

We have applied a field theoretic treatment to the vector meson sector, and
in particular, the $\rho$.  Starting with the HLS model allows one to
produce a Lagrangian for the low energy hadronic sector, which, when
carefully treated is seen to be consistent with physical expectations.
Working with this model, we have demonstrated the automatic preservation of
the massless photon and the equivalence of VMD1$_a$ and VMD2$_a$ (the HLS
generalizations of VMD1 and VMD2).  Previous fits of the pion form factor,
preferred VMD1$_a$ relative to the standard VMD1, which may have further
phenomenological consequences.  Finally, as predicted from S-matrix theory,
we have shown that the determination of the complex pole position of the
$\rho$ meson from a large number of existing fits to data is model and
process independent.  This will be of particular use for the comparison of
$\rho$ parameter determinations from different experiments and for
exploring the usefulness and possible limitations of different realizations
of vector meson dominance.

\vspace{1.0cm}
\begin{center}
{\bf Acknowledgements}
\end{center}
We would like to thank S.~Gardner,
K.F.~Liu, J.A.~Oller, J.R.~Palaez, M.J.~Peardon and J.~Sloan
for helpful correspondence and discussions.
This work is supported by the
US Department of Energy under grant DE--FG02--96ER40989 (HOC)
and the Australian Research Council (AGW).

\begin{table}[htb]

\caption{
\label{one}
}

\nc{\first}{^{+3.7}_{-2.8}}
\nc{\second}{^{+2.1}_{-1.9}}
\nc{\third}{^{+4.9}_{-2.8}}
\nc{\fourth}{^{+1.8}_{-2.0}}
\nc{\fifth}{^{+1.8}_{-1.5}}
\nc{\sixth}{^{+3.2}_{-2.9}}

\begin{tabular}{lccccc}
Fit & $\chi^2/$dof & $m_\rho$ (MeV) & $\Gamma_\rho$ (MeV)
& $\overline{m}_\rho$ (MeV) & $\overline{\Gamma}_\rho$ (MeV) \\
\hline
I  &129/133&$775.9\pm1.1$&$150.5\pm3.0$& $758.3$     & $145.0$      \\
IIA& 68/75 &$763.1\pm3.9$&$153.8\pm1.2$&$756.3\pm1.2$& $141.9\pm3.1$\\
IIB& 66/76 &$771.3\pm1.3$&$156.2\pm0.4$&$757.0\pm1.0$& $141.7\pm3.0$\\
IIC& 67/76 &$773.9\pm1.2$&$157.0\pm0.4$&$757.0\pm1.0$& $141.7\pm3.0$\\
IID& 68/76 &$773.9\pm1.2$&$146.9\pm3.4$&$757.0\pm1.0$& $141.7\pm3.0$\\
III&       &             &             &$757.5\pm1.5$& $142.5\pm3.5$\\
IV[VMD1$_a$]&65/77&$774.7\pm0.7$&$147.1\pm1.6$&$758.5$&$142.8  $ \\
IV[VMD2]&81/77&$780.4\pm0.7$&$155.4\pm2.0$&$760.0$&$146.8      $ \\
IV[HLS(VMD2$_a$)]&65/77&$775.2\pm0.7$&$147.7\pm1.5$&$759.0$&$143.4$ \\
IV[WCCWZ]&61/76&$770.9\fifth$&$140.6\sixth$&$759.2$&$139.6$ \\
V&       &   771         &  147          &$756$         & $142$  \\
VI[KS]&81/65& $774.9\pm0.9$ & $144.2\pm1.5$ & $759.1$& $139.9$   \\
VI[GS]&54/65& $776.4\pm0.9$ & $150.5\pm1.6$ & $758.8$& $145.0$   \\
\end{tabular}
\end{table}

Table \ref{one} shows the extraction of $\rho$ masses and widths for both
the real 
part of the complex
pole ($\overline{m}_\rho$ and $\overline{\Gamma}_\rho$
defined in Eq.~(\ref{pole_def})) and 
the
standard real axis ($m_\rho$ and $\Gamma_\rho$ as given in 
Eq.~(\ref{usual}))
prescriptions. Fits I (Ref.~\cite{barkov}),
II (Refs.~\protect{\cite{GO,GO2}}), III (Ref.~\cite{BCP}) and
IV (the results of the unitarized fits of Ref.~\cite{BEMOSW},
see Table 2 therein)
use the \ep data obtained in I.
Fit V (Ref.~\cite{scatt2},
adjusted to $\overline{m}$ from $\widetilde{m}$) uses $\pi\pi$ scattering
data
and Fit VI (Ref.~\cite{aleph} using two different fitting functions,
see Table 3 therein)
gives results for the charged $\rho$ obtained from $\tau$ decay.

\end{document}